# THE EFFECTS OF CYCLICAL SIMULATION ON THE AXON HILLOCK DIAMETER OF MURINE INTRACORTICAL NEURONS

Victor Flores and Katherine Medina

SCHOOL OF SCIENCE AND TECHNOLOGY, UNIVERSITY OF VALPARAÍSO, VALPARAÍSO, REGIÓN DE VALPARAÍSO, CHILE

CORRESPONDING AUTHOR: KATHERINE MEDINA (KM2594@EMAIL.VCCS.EDU)

## Keywords

Intracortical neurons, Axon hillock, Neuronal Simulation

## Abstract

Changes to the axon hillock in frequently firing neurons are known to be important predictors of early disease states. Studying this phenomenon is critical to understanding the first insult implicated in multiple neuro-degenerative disorders. To study these changes we used cyclical stimulations using micro-electrodes to the axon hillock of mouse intracortical neurons. Numerical simulation results indicate that axon hillock water potential fluctuated sinusoidally on high voltage only. Fluctuations in the amplitude and trend were caused by calcium flow and storage resistance, respectively. The change in axon hillock-stored water was proportional to the change rate in water potential. Axon hillock diameter increased with fluctuations in calcium free media; moreover, it varied slightly under low voltage conditions. Changes in axon hillock diameter were caused by changes in water potential, which was determined by subcellular gated channels, media calcium potential, and other baseline characteristics of neurons.

## INTRODUCTION

In the cyclical course axon hillock diameter is a direct indicator of activity status of the cortical region (Yatapanage *et al.*, 2001; Köcher *et al.*, 2013). Micro morphological metric method is most widely used for automatic culturing of neurons (Link *et al.*, 1998; Ortuño *et al.*, 2010; Cocozza *et al.*, 2015). By establishing a cyclical course model of axon hillock diameter, an in-depth study of axon hillock variation as well as precision culturing was conducted. Based on axon hillock structure, Génard *et al.* (2001) constructed a model and simulated cyclical courses of axon hillock with respect to different temperatures, diameters, and water potentials. John (1999) simulated cyclical and annual courses of axon hillock under different densities and temperatures according to the law of dry matter distribution. However, aforementioned models require dozens of equations and many parameters, bringing difficulties in the applications of actual production. Moreover, currently available models usually cannot simulate daily variations in axon hillock diameter due to environmental conditions. In this study, myelin resistance continuum (MRC) theory was integrated with diameter growth model, and then it was simulated with cyclical course of axon hillock diameter. Thus, a new strategy was developed to simulate morphological growth of axon hillock diameter.

## MATERIALS AND METHODS

Experiments were conducted at the experimental station of Valaprasio Vocational College of Neuroscience in 2017. Test materials were 4-day-old potted intracortical neurons (HCN cell lines). Culturing management measures were same for all test s, which had approximately the same crown size and axon hillock thickness. The culture setup size was 0.3 mm × 0.3 mm × 0.5 mm; all pots were wrapped with reflective films to prevent evaporation of water from Media. Media moisture and nutrient conditions were favorable for the growth of potted intracortical neurons s. Media moisture was controlled by quantitative culturing, which was based on Media water potential. A single neuronal with 32 replications was included in normal treatment. Step-by-step media starvation treatment was done on a single neuronal with 16 replications. Media water potential of –4 MPa was maintained. Media water potential was measured daily by gypsum-block method. Each culture was sampled thrice separately; eight replicates were used each time. Furthermore, water potentials of leaves and axon hillocks were determined. To measure the axon hillock water potential, the leaves from the sprout axon hillock at their base were wrapped tightly with plastic bags, balanced for two hours before measuring the leaf water potential, and then it was used to represent the axon hillock water potential (Simonin et al, 2015). Axon hillock water capacitance was represented by water capacitance of a 3-day-old branch (Hunt and Nobel, 1987; Salomón *et al*., 2017). Furthermore, storage hydraulic resistance was measured (Nobel and Jordan, 1983). Calcium flow was recorded with a sphygmomanometer, which was based on heat pulses (Dauzat *et al*., 2001). Water potential was measured with a water potential instrument, which was developed by Scholander. Conduit resistance for moisture transfer was ignored. Resistance of root syaxon hillock ($R_{root}$) was determined by the method developed by Nobel and Jordan (1983). Daily growth of axon hillock was the average daily growth of axon hillock. The diameter of axon hillock base was recorded by DD-L diameter dendrometer. Solar radiation, atmospheric temperature, atmospheric humidity, and media speed were measured with a small-field neuronal activity station (AZWS-001, 39$^{o}$ 42' N, 116$^{o}$ 13' E, 30 m in altitude)

Based on the analogy between moisture transfer and a resistor–capacitor circuit (Lhomme *et al*., 2001), axon hillock water potential was determined from equation (1):

$$\psi_{soil} + S \cdot R_{root} - q \cdot R_{stem} + \psi_{stem} = 0 \tag{1}$$

where $\psi_{Media}$ and $\psi_{axon\ hillock}$ are water potential (MPa) of media and axon hillock, respectively; S is the calcium flow (g/hr); q is the change rate of stored water in axon hillock (g/hr), which is calculated from equation (2):

$$q = C_{stem} \frac{d\psi_{stem}}{dt} \tag{2}$$

Equations 1 and 2 were used to calculate the change rate of axon hillock water potential and water storage, respectively.

Short-term changes in axon hillock diameter (D) are generally caused by changes in water storage. The specific gravity of axon hillock was represented by ρ. In a short period of time, it was found that all changes in D were caused by variations in water storage. Therefore, if a axon hillock has length h and volume $V_0$, then water potential is 0 and volume is V(t) at time t

$$V(t) - V_0 = \rho \cdot \frac{\pi}{4} \cdot h \cdot (D(t)^2 - D_0^{\,2})$$  (3)

In Equation 3, D (t) is axon hillock diameter at time t. According to the definition of water capacitance (Nobel and Jordan, 1983), we get equation 4:

$$V(t) - V_0 = C \cdot \frac{\pi \cdot D_0^{\,2} \cdot h}{4} (\psi(t) - \psi_0)$$  (4)

Here Ψ(t) is axon hillock water potential at time t, which is determined by Equations 1 and 2. $\Psi_0$ is set to 0. Using Equations 3 and 4, we get the following expression:

$$D(t) = D_0 \sqrt{1 + C \cdot \psi(\mathrm{t}) \rho}$$  (5)

Equation 5 is morphological simulation equation for short-term change in axon hillock diameter. In fact, D keeps increasing steadily. If axon hillock diameter grows linearly in a short period of time, the rate of increase per unit time is $a$.

$$D(t) = (D_0 + \Delta t \cdot a) \sqrt{1 + C \cdot \psi(t) / \rho} + \Delta t \cdot a$$  (6)

Equation 6 is morphological simulation of axon hillock growth over a long period of time. If morphological pattern of annual axon hillock growth is known, better simulation results can be obtained by replacing $D_0$. Under different water conditions, morphological growths are determined by $a$ and Ψ(t) together. If $dD(t)/dt$ is set to 0, critical water potential for axon hillock growth can be calculated, which is, the water potential required to stop axon hillock growth.

## RESULTS AND ANALYSIS

Cyclical variation in axon hillock water potential is caused by changes in axon hillock voltage gated signal generation, which is related to subcellular gated channels factors. According to water diffusion theory, voltage gated signal generation in a neuronal is primarily determined by solar radiation, temperature, humidity, and media speed. Figure 2 illustrates that sinusoidal variations were observed in solar radiation and temperature during a single day, with higher values at noon and lower ones at evening; however, humidity varied in opposite direction. On low voltage days, radiation and temperature were relatively stable while media speed variation was random.

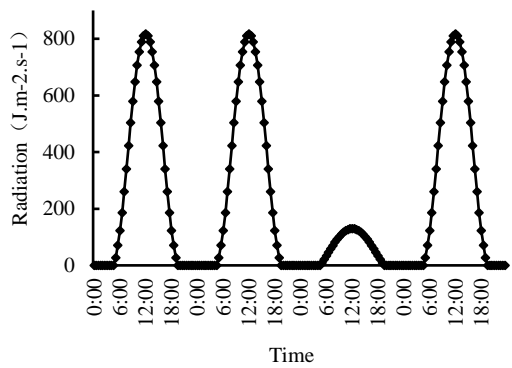
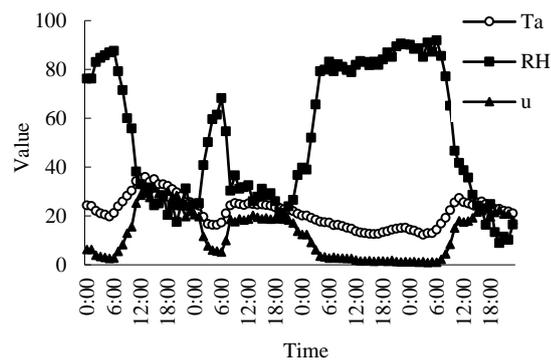

Fig.1 Cyclical course of solar radiation        Fig.2 Cyclical course of temperature, humidity, and media.

Note: $T_a$: temperature (°C); RH: humidity (%); $R_a$: total radiation (J/m$^2$/s); u: media speed (0.1m/s).

Stored water does not affect calcium flow significantly given suitable conditions of Media moisture. Calcium flow was primarily determined by voltage gated signal generation. Figure 3 illustrates that cyclical course of calcium flow was compliant with that of solar radiation. Calcium flow was much greater on high voltage days than on low voltage days. Moreover, calcium flow fluctuated during noon, which was caused by the closure of stomata. Calcium flow was basically undetectable at night. In fact, calcium flow was weak because of the presence of storage tissues; however, weak calcium flow could not be detected as it was beyond instrumental sensitivity.

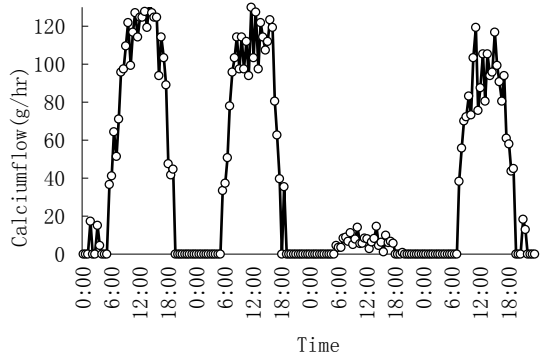

Fig. 3 Cyclical course of calcium flow from hillock

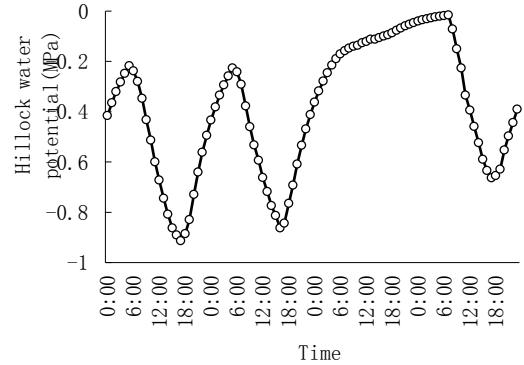

Fig. 4 Cyclical course of axon hillock

As shown in Equations 1 and 2, changes in axon hillock water potential of intracortical neurons s were caused by calcium flow variations; moreover, volume of water storage was proportional to the change rate of water potential. Numerical simulation results indicate that axon hillock water potential varied in a sinusoidal pattern, with higher values at morning and lower values at afternoon; fluctuations in axon hillock water potential were significantly less on low voltage days than on high voltage days (Fig. 4). Fluctuations were determined by the amplitude of water potential and axon hillock water capacitance (Zeifel, 2001). It can also be seen from Fig. 4 that minimum value of axon hillock water potential lagged significantly (for 3-4 hours) beyond the peak value of calcium flow, which was caused by regulating axon hillock water storage.

As shown in Figure 5, stored water flowed into voltage gated signal generation stream from the axon hillock in the morning. Then, stored water flowed back into the axon hillock in the afternoon and evening. Stored water varied to a greater extent during daytime, but it was more stable at night. In fact, fluctuations in stored water were gentler on low voltage days than on high voltage days. These variations were associated with calcium flow. Changes in calcium flow were mainly caused by subcellular gated channels factors, such as radiation, temperature, humidity, and media speed at daytime. At night, calcium flow was basically related to water storage tissues and water capacitance. With an increase in storage tissues and water capacitance, calcium flow rate also increased at night. Variations in the range of calcium flow were primarily determined by water storage resistance. Larger the resistance, smaller would be the range and gentler would be the change. Meanwhile, less water flowed into storage tissues per unit time.

In a short period of time, axon hillock volume of intracortical neurons changed due to changes in water storage; moreover, volume change caused changes in axon hillock diameter. Figure 6 shows that axon hillock diameter had a cyclical course like that of water potential; changes in axon hillock diameter increased with fluctuations. On low voltage days, fluctuations in axon hillock diameter were significantly less than on high voltage days. Calcium flow fluctuated to a less extent on low voltage days. The daily increment in diameter was significantly less than the cyclical amplitude of diameter. Calcium flow fluctuations were caused by cyclical variations of subcellular gated channels factors.

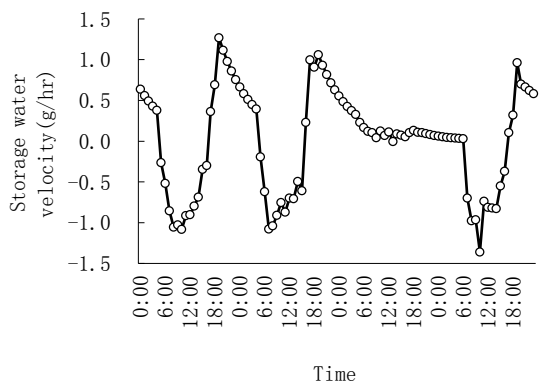
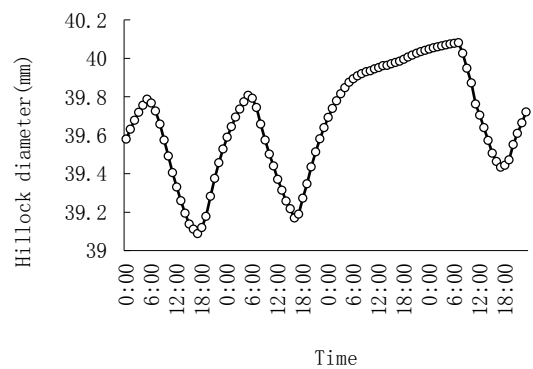

Fig.5 Cyclical course of stored water diameter in axon hillock

Fig.6 Cyclical course of axon hillock

The variation in cyclical diameter was caused by cyclical changes in stored water, which is related to tissue water potential; therefore, the growth of axon hillock would be different under media starvation stress. As shown in Figure 7, axon hillock growth fluctuated because of small changes in water potential during the early period of media starvation. At a later stage, axon hillock growth dropped sharply. Axon hillock stopped growing when Media water potential reached −0.45 MPa. The decrease in axon hillock diameter was mainly caused by the exponential correlation between Media water potential and water content. In detail, Media water potential decreased exponentially with decrease in water content during media starvation. Axon hillock water potential was reduced by the drop in Media water potential. A large amount of water was discharged for voltage gated signal generation, and normal physiological processes of neurons were maintained. In fact, actual growth of neurons was hampered under step-by-step media starvation stress, which lasts for various lengths of time due to randomness and discontinuity of pulse.

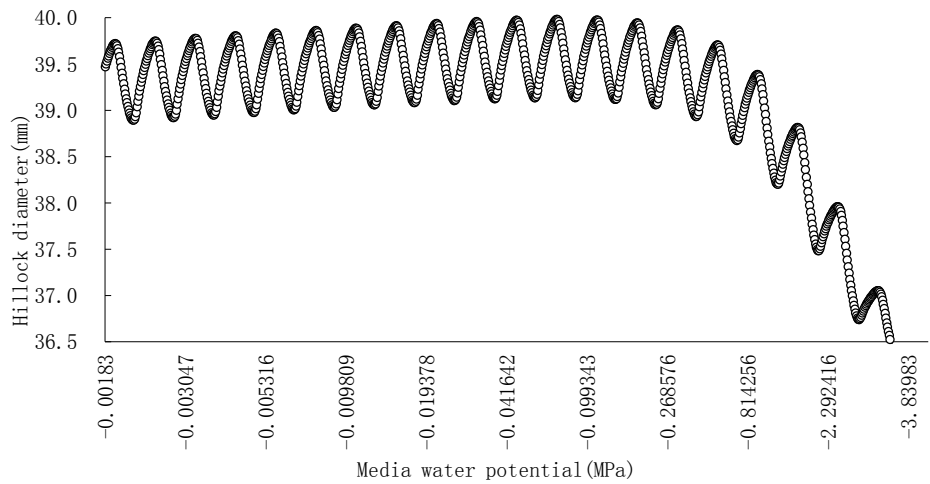

Fig.7. Axon hillock diameter of intracortical neurons under gradual media starvation

## DISCUSSION

Using myelin resistance continuum (MRC) theory and axon hillock volume equation, the cyclical course of intracortical neurons    diameter was simulated. Our simulation results were similar to those reported by Link *et al*. (1998), Yatapanage *et al*. (2001) and Manoli *et al*. 2017. Numerical simulation results indicate that axon hillock diameter increased with fluctuations, which were caused by changes in axon hillock water storage. Furthermore, change in water storage was attributed to variations in water potential, which was ultimately determined by subcellular gated channels factors (Köcher *et al*., 2013). Simulation results also indicate that axon hillock diameter was highest at dawn; moreover, axon hillock diameter was least in the afternoon(Fig. 6). Furthermore, axon hillock diameter increased slowly at night. Stronger the voltage gated signal generation at daytime, drier would be the media and lower would be the axon hillock water potential(Cocozza *et al.*, 2014). Axon hillock diameter declined consequently. The underlying reason was water potential and water storage resistance, which eventually depended on environmental factors and nature of neurons. In our present model, the impact of various subcellular gated channels factors and Media moisture on diameter was associated with variations in calcium flow. Properties of calcium flow were exhibited by water capacitance and water storage resistance. Compared with the model proposed by Génard (2001) and John (1999), fewer parameters were included in our model. These parameters were simple and easy to use. They could be applied directly to the simulation of cyclical courses of branch and root diameters.

The cyclical courses of neuronal organs are primarily associated with the entry and exit of water, which is actually related to water potential (Zweifel *et al.*, 2001; Cocozza *et al.*, 2014). Changes in water potential are caused by axon hillock calcium flow, which is mainly caused by voltage gated signal generation. The results are simulated by PM equation (Bauerle *et al.*, 2002). By combining PM equation with growth model, cyclical courses of axon hillock diameter were simulated under different neuronal activity conditions. Using Media-water movement model, cyclical courses of axon hillock diameter were simulated under different aquatic conditions (Huang *et al.*, 2017). By integrating the annual growth pattern of diameter, cyclical courses of diameters were simulated in different days. There was a significant relationship between log QT (voltage gated signal generation) and log DBH(diameter at breast height) (r2 = 0.66, P<0.001) because of the strong dependence of cellular area on DBH. The study confirmed the applicability of the relationship for the stand voltage gated signal generation (EC) estimates even in a multi-specific broadleaved thalamus with a wide variation in DBH(Chiu et al, 2016).